\documentclass[reprint,
 amsmath,amssymb,
 aps,
]{revtex4-2}

\usepackage{graphicx}% Include figure files
\usepackage{dcolumn}% Align table columns on decimal point
\usepackage{bm}% bold math
\usepackage{xcolor}
\usepackage{xcolor, soul}
\sethlcolor{cyan}
\begin{document}

\preprint{APS/123-QED}

\title{Axial spectral encoding of space-time wave packets}

\author{Alyssa M. Allende Motz}
\author{Murat Yessenov}
\author{Ayman F. Abouraddy}
\affiliation{CREOL, The College of Optics \& Photonics, University of Central Florida, Orlando, Fl 32816, USA}

\email{raddy@creol.ucf.edu}

% To be edited by editor
% \dates{Compiled \today}

% To be edited by editor
% \doi{\url{http://dx.doi.org/10.1364/optica.XX.XXXXXX}}

\begin{abstract}
Space-time (ST) wave packets are propagation-invariant pulsed optical beams whose group velocity can be tuned in free space by tailoring their spatio-temporal spectral structure. To date, efforts on synthesizing ST wave packets have striven to maintain their propagation invariance. Here, we demonstrate that one degree of freedom of a ST wave packet -- its on-axis spectrum -- can be isolated and purposefully controlled independently of the others. Judicious spatio-temporal spectral amplitude and phase modulation yields ST wave packets with programmable spectral changes along the propagation axis; including red-shifting or blue-shifting spectra, or more sophisticated axial spectral encoding including bidirectional spectral shifts and accelerating spectra. In all cases, the spectral shift can be larger than the initial on-axis bandwidth, while preserving the propagation-invariance of the other degrees of freedom, including the wave packet spatio-temporal profile. These results may be useful in range-finding in microscopy or remote sensing via spectral stamping.
\end{abstract}

\maketitle

\section{Introduction}

The unique opportunities offered by jointly manipulating the spatial and temporal degrees of freedom of an optical field -- rather than modulating the field separately in space \cite{GoodmanBook05} or time \cite{Weiner09Book} -- have started recently to capture considerable attention \cite{Kondakci16OE,Parker16OE,Wong17ACSP1,Wong17ACSP2,Porras17OL,Efremidis17OL,Shaltout19Science,Chong20NP}. One example within this general enterprise is that of `space-time' (ST) wave packets, which are a class of pulsed beams endowed with a spatio-temporal structure in which each spatial frequency is associated with a single wavelength \cite{Donnelly93PRSLA,Kondakci17NP,Yessenov19OPN}, by virtue of which they exhibit unique characteristics; e.g., propagation-invariance \cite{Porras17OL,Kondakci17NP,PorrasPRA18,Bhaduri18OE,Bhaduri19OL,Schepler20arxiv,Shiri20arxiv}, controllable group velocities \cite{Wong17ACSP2,Kondakci19NC,Bhaduri19Optica,Yessenov19OE}, self-healing \cite{Kondakci18OL}, and omni-resonant interaction with planar cavities \cite{Shiri20OL,Shiri20APLP}. Although the study of propagation-invariant pulsed beams extends for almost 40~years \cite{Brittingham83JAP,Ziolkowski85JMP,Saari97PRL,Kiselev07OS,Turunen10PO,FigueroaBook14}, the focus has always been on maintaining the field invariant rather than controlling the axial variation of one characteristic while maintaining the others fixed. We pose here the following question: can a ST wave packet remain propagation-invariant while controlling the axial evolution of \textit{one} of its aspects in isolation from all the other characteristics of the wave packet? For example, can we control the evolution of the on-axis spectrum of a ST wave packet \textit{without} impacting its propagation invariance or the tunability of its group velocity? This would produce a \textit{continuous, axial spectral encoding} that can be useful in range-finding through spectral stamping of reflections from objects at different distances -- whether in the context of
optical microscopy or remote sensing.

Recently, an alternative approach to tuning the group velocity of an optical wave packet known as the `flying-focus', which does \textit{not} provide diffraction-free behaviour, was proposed theoretically \cite{SaintMarie17Optica} and demonstrated experimentally \cite{Froula18NP,Jolly20OE}. This approach relies on longitudinal chromatism in which a temporally chirped pulse is focused by a lens having prescribed chromatic aberrations. In this strategy, the on-axis spectrum shifts continuously with propagation, but this shift is \textit{not} independent of the group velocity; in fact, the rate of the spectral shift dictates the group velocity of the flying-focus. In other words, the group velocity and the axial evolution of the spectrum are inextricably linked and cannot be manipulated independently. This helps sharpen the question posed above: can we control the spectral evolution and the group velocity of a wave packet \textit{independently} through sculpting its spatio-temporal structure? More generally, can we produce an arbitrary axial spectral encoding for an optical wave packet while maintaining its other desirable characteristics (e.g., propagation-invariance) intact? 

Here we show that ST wave packets can support controllable and readily adaptable evolution of the on-axis spectrum along the propagation direction. Such axial spectral encoding includes red-shifting or blue-shifting spectra, complex bidirectional-shifting spectra in which the direction of the spectral shift can switch at prescribed axial locations, and accelerating spectra. This control can be exercised independently of the wave-packet group velocity (whether subluminal or superluminal), and without affecting its propagation invariance. Time-resolved measurements of the ST wave packet confirm that its spatio-temporal profile remains invariant with propagation despite the underlying spectral changes. Our approach is independent, in principle, of the selected wavelength \cite{Yessenov20OSAC} or bandwidth \cite{Kondakci18OE} of the field. These features are achieved through sculpting the amplitude \textit{and} phase distribution imparted to the spectrally resolved wave front, in contrast to our previous work that made use of purely \textit{phase} modulation to synthesize propagation-invariant ST wave packets \cite{Yessenov19OPN}. Our results highlight a crucial aspect of modulating the joint spatio-temporal spectrum rather than manipulating the spatial (Fourier optics \cite{GoodmanBook05}) or the temporal (pulse shaping \cite{Weiner09Book}) degrees of freedom separately. Indeed, whereas spectral \textit{phase} modulation is a staple of ultrafast pulse shaping, spectral \textit{amplitude} modulation would result in irrecoverable spectral filtering, and is hence usually avoided. Uniquely, joint spatio-temporal spectral amplitude modulation can be utilized \textit{without} necessarily filtering or eliminating any portion of the temporal spectrum. This strategy opens up a plethora of opportunities for programmable axial spatial encoding of ST wave packets -- independently of the wave-packet group velocity, extending from range-finding in microscopy and remote sensing, to nonlinear interactions in plasmas \cite{Turnbull18PRL1,Turnbull18PRL2}.

\section{Space-time wave packets}

\begin{figure}[ht!]
\centering
\includegraphics[width=8.6cm]{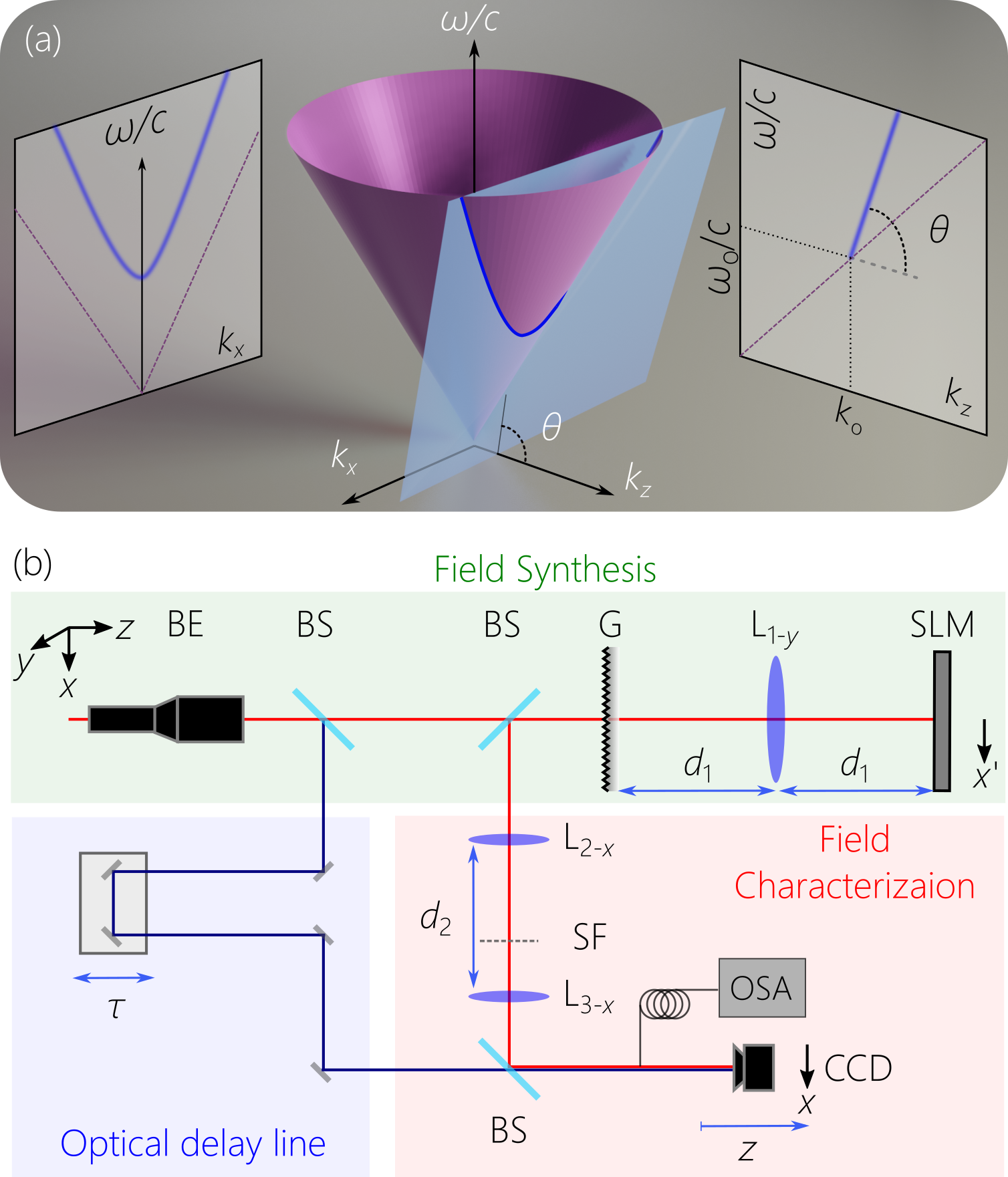}
\caption{(a) The spatio-temporal spectral support domain of a ST wave packet at the intersection (blue curve) of the light-cone $k_{x}^{2}+k_{z}^{2}\!=\!(\tfrac{\omega}{c})^{2}$ with a spectral plane tilted by angle $\theta$ with respect to the $k_z$-axis. Also shown are the spectral projections onto the $(k_{z},\tfrac{\omega}{c})$ and $(k_{x},\tfrac{\omega}{c})$ planes, which are a straight line and a conic section, respectively. (b) Schematic of the optical setup to synthesize and characterize ST wave packets endowed with prescribed axial spectral encoding. BE: Beam expander; BS: beam splitter; G: diffraction grating; SLM: spatial light modulator; SF: spatial filter; OSA: optical spectrum analyzer; L: cylindrical lens. The distance $d_{1}\!=\!500$~mm is equal to the focal length of L$_{1-y}$, and $d_{2}\!=\!500$~mm is equal to the sum of the focal lengths of L$_{2-x}$ (100~mm) and L$_{3-x}$ (400~mm). The transverse coordinate at the SLM plane is $x'$ and at the CCD is $x$.}
\label{fig:schematic}
\end{figure}

The concept of propagation-invariant ST wave packets is by now well-established \cite{Kondakci17NP,Yessenov19OE,Yessenov19OPN}, and we provide here only a brief overview to lay the groundwork for the subsequent formulation of axial spectral encoding. Such a wave packet in free space is a pulsed optical beam whose spatio-temporal spectral support domain lies along the intersection of the light-cone $k_{x}^{2}+k_{z}^{2}\!=\!(\tfrac{\omega}{c})^{2}$ with a tilted spectral plane $\tfrac{\omega}{c}\!=\!k_{\mathrm{o}}+(k_{z}-k_{\mathrm{o}})\tan{\theta}$; where $k_{x}$ and $k_{z}$ are the transverse and longitudinal components of the wave vector along $x$ and $z$, respectively, $\omega$ is the temporal (angular) frequency, $c$ is the speed of light in vacuum, $\omega_{\mathrm{o}}$ is a fixed temporal frequency, $k_{\mathrm{o}}\!=\!\tfrac{\omega_{\mathrm{o}}}{c}$ is the corresponding wave number, and $\theta$ is the angle made by the spectral plane with respect to the $k_{z}$-axis [Fig.~\ref{fig:schematic}(a)]. We refer to $\theta$ as the spectral tilt angle and to $k_{x}$ as the spatial frequency. Such a field $E(x,z;t)\!=\!e^{i(k_{\mathrm{o}}z-\omega_{\mathrm{o}}t)}\psi(x,z;t)$ can be shown to have a propagation-invariant envelope $\psi(x,z;t)\!=\!\psi(x,0;t-z/\widetilde{v})$ that travels rigidly in free space at a group velocity $\widetilde{v}\!=\!c\tan{\theta}$ independently of beam size, pulse width, or central wavelength -- as demonstrated recently \cite{Kondakci19NC}. Indeed, the spectral projection onto the $(k_{z},\tfrac{\omega}{c})$-plane is a straight line, indicating absence of group velocity dispersion [Fig.~\ref{fig:schematic}(a)]. The projection onto the $(k_{x},\tfrac{\omega}{c})$-plane is a conic section \cite{Yessenov19PRA} that can be approximated by a parabola for narrow temporal bandwidths $\Delta\omega\!\ll\!\omega_{\mathrm{o}}$, indicating that each spatial frequency is associated with a single temporal frequency [Fig.~\ref{fig:schematic}(a)]:
\begin{equation}\label{Eq:kxOmega}
k_{x}(\omega;\theta)=\frac{1}{c}\sqrt{2\omega_{\mathrm{o}}(\omega-\omega_{\mathrm{o}})(1-\cot{\theta})}.
\end{equation}
Tuning the group velocity requires changing the \textit{curvature} of this spatio-temporal spectral parabola, which can be readily achieved via phase-only spectral modulation \cite{Kondakci19NC,Bhaduri19Optica,Yessenov19OPN}.

\begin{figure*}[ht!]
\centering
\includegraphics[width=11.5cm]{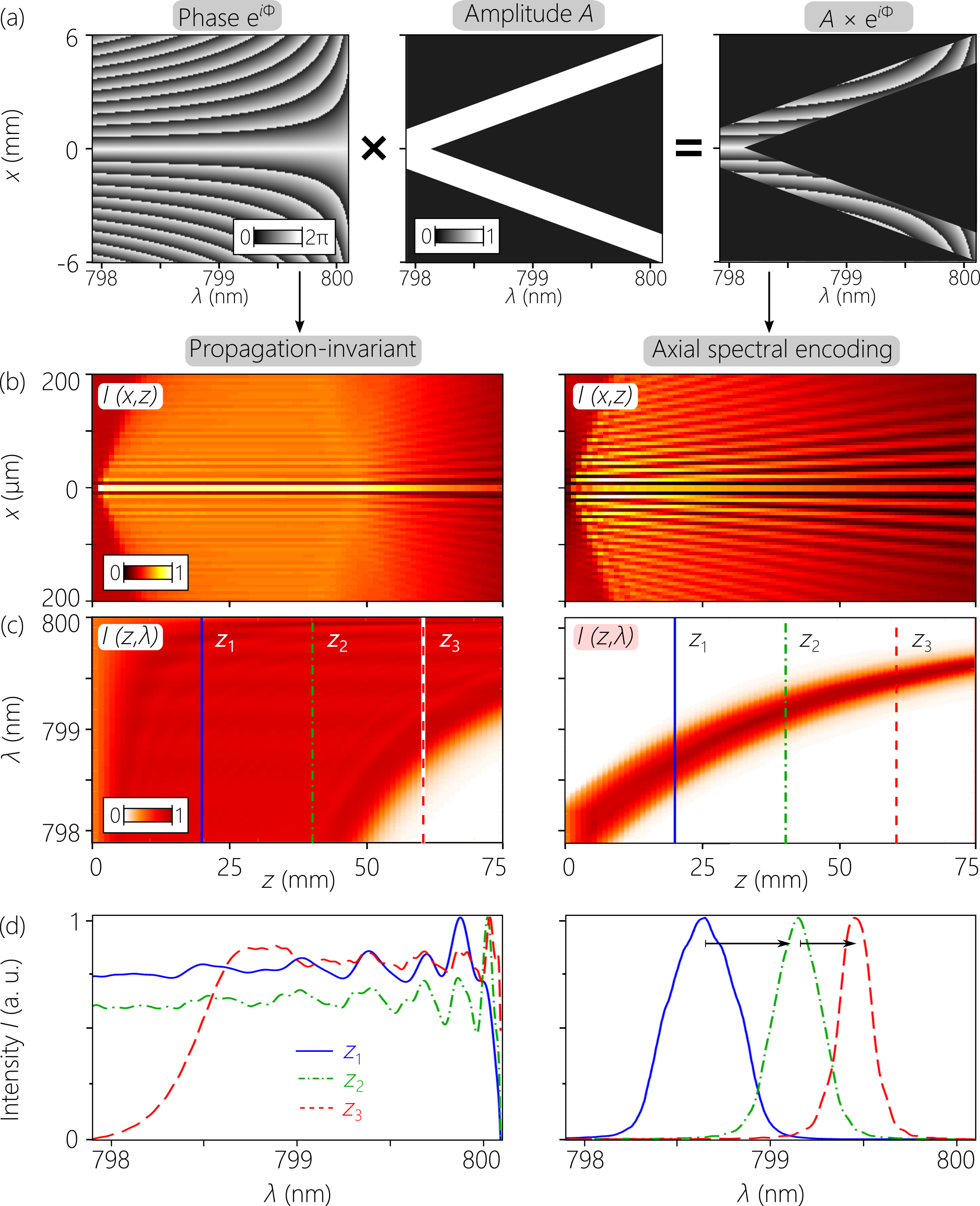}
\caption{(a) Axial spectral encoding through phase \textit{and} amplitude spectral modulation. A typical phase distribution $\Phi$ (left) for the synthesis of a propagation-invariant ST wave packet ($\theta\!=\!50^{\circ}$ and $\widetilde{v}\!=\!c\tan{50^{\circ}}\!\approx\!1.19c$) is combined with an amplitude distribution $A$ (middle) to produce an amplitude-masked phase distribution $Ae^{i\Phi}$ (right), which yields a ST wave packet with tailored axial spectral encoding but the \textit{same} group velocity. The phase \textit{and} amplitude masks are realized by the SLM in conjunction with a spatial filter that rejects the non-diffracted light}. (b) Calculated time-averaged intensity $I(x,z)$ for propagation-invariant (left) and axial spectral encoded (right) ST wave packets. Note the very different axial and transverse spatial scales. (c) Calculated spectral evolution $I(z,\lambda)$ for the ST wave packets in (b). (d) Spectra at selected axial planes for the ST wave packets in (b). The spectrum of the propagation-invariant ST wave packet (left) does not change significantly with propagation, whereas that of the axially spectral-encoded ST wave packet (right) undergoes significant spectral evolution with propagation.
\label{fig:theory}
\end{figure*}

Such a ST wave packet can be prepared via the two-dimensional pulse synthesizer depicted in Fig.~\ref{fig:schematic}(b) starting with femtosecond pulses from a Ti:sapphire laser (Tsunami, Spectra Physics) having a central wavelength $\lambda_{\mathrm{o}}\!=\!799$~nm and a pulse duration of $\sim120$~fs (as verified via a FROG measurement \cite{DeLong1994frequency} obtained by a GRENOUILLE \cite{Akturk2003OE} commercial platform). The spectrum of the plane-wave (30-mm-diameter) pulses are spread in space by a diffraction grating (1200~lines/mm; Newport 33009FL01-360R) and collimated by a cylindrical lens before impinging on a spatial light modulator (SLM; Hamamatsu X10468-02). The SLM imparts a two-dimensional phase distribution $\Phi$ to the spatially resolved spectrum, which is then retro-reflected through the lens and reconstituted by the grating to produce the ST wave packet [Fig.~\ref{fig:schematic}(b)]. The SLM-imparted phase distribution [Fig.~\ref{fig:theory}(a)] can be designed to implement any spectral tilt angle $\theta$ by assigning to each wavelength $\lambda$ incident on the SLM a phase distribution corresponding to a particular spatial frequency $k_{x}(\omega;\theta)$ according to Eq.~\ref{Eq:kxOmega} \cite{Kondakci17NP,Yessenov19OPN}.  In all cases, a spatial filter (in the form of a low-pass rejection filter to eliminate the spatial components in the immediate vicinity of $k_{x}\!=\!0$) is placed in the Fourier plane between two cylindrical lenses with their focusing axis parallel to the $x$-coordinate, to remove any non-diffracted portion of the field (due to the finite diffraction efficiency of the SLM).

We characterize the synthesized ST wave packets in three domains [Fig.~\ref{fig:schematic}(b)]. (1) A CCD camera (TheImagingSource DMK 33UX178) is scanned along the $z$-axis in the path of the ST wave packet to record the time-averaged intensity $I(x,z)$. (2) A 400-$\mu$m-diameter multimode fiber (Thorlabs, M74L01-$\emptyset400$~$\mu$m, 0.39~NA), connected to an optical spectrum analyzer (OSA; Advantest q8381, 0.1-nm-resolution), is scanned along the $z$-axis to record the axial spectral evolution of the ST wave packet $I(z,\lambda)$. A multimode is selected to increase the signal-to-noise ratio and thus confirm the predicted effect unambiguously. The diameter was selected to capture a large portion of the field energy around the optical axis and thus avoid any spurious spectral effects that can arise if a single-mode fiber is used instead. The measured spectrum is thus representative of the axial field evolution. However, the multimode fiber introduces $\approx30\%$ spectral broadening compared to a single-mode fiber, and the results reported here are therefore a conservative underestimate of the axial spectral encoding offered by our approach. (3) The spatio-temporal profile of the ST wave packet is measured by interfering the wave packet with the original Ti:sapphire laser pulses $I(x,\tau)$ at any selected axial plane $z$. By placing the pulse synthesizer in one arm of a Mach-Zehnder interferometer, and placing an optical delay in the reference arm traversed by the femtosecond pulses, time-resolved measurements are obtained from the visibility of the spatially resolved interference fringes that are observed when the ST wave packet and the reference pulse overlap in space and time. This arrangement is also used to estimate the group velocity $\widetilde{v}$ of the ST wave packet \cite{Kondakci19NC,Bhaduri19Optica,Yessenov19OE}.

\section{Space-time wave packets with axial spectral encoding}

An example of the phase distribution $\Phi$ imparted by the SLM to realize a ST wave packet is shown in Fig.~\ref{fig:theory}(a), corresponding to a spectral tilt angle $\theta\!=\!50^{\circ}$ (superluminal group velocity $\widetilde{v}\!\approx\!1.19c$) \cite{Kondakci17NP}. Implementing such a phase distribution results in a diffraction-free beam [Fig.~\ref{fig:theory}(b)] whose spectrum is invariant [Fig.~\ref{fig:theory}(c-d)] over its propagation length. We proceed to show that the key to controlling the axial spectral evolution of the wave packet is modulating the \textit{amplitude} of its spatio-temporal spectrum at the SLM plane -- a degree of freedom that has hitherto gone unexploited. 

Note that the amplitude modulation is \textit{not} introduced by means of an additional device or component. Rather, the SLM serves as a phase and amplitude modulator. The continuous sections of the SLM where the phase is set to zero produce a field whose spatial spectrum is in the vicinity of $k_{x}\!=\!0$, which is then eliminated by the spatial filter (SF) along with the non-diffracted field from the SLM.

Consider the configuration shown in Fig.~\ref{fig:theory}(a) where we combine the phase distribution $\Phi$ (corresponding to a particular spatial tilt angle $\theta$) with a two-dimensional amplitude masking distribution $A$. Consequently, the phase modulation imparted by the SLM is now restricted to a particular portion of the spectrally spread wave front as dictated by the amplitude mask, $Ae^{i\Phi}$. Experimentally, this amplitude modulation is achieved by setting the SLM phase to zero wherever the target amplitude is required to be zero, and then subsequently inserting a spatial filter in the Fourier plane between the cylindrical lenses L$_{2-x}$ and L$_{3-x}$ placed in the path of the ST wave packet [Fig.~\ref{fig:schematic}(b)] to eliminate the unscattered (or zero-phase) field component.

From Fig.~\ref{fig:theory}(a) we can understand why \textit{joint} spatio-temporal spectral amplitude modulation does \textit{not} necessarily eliminate a wavelength altogether as it would in a traditional ultrafast pulse shaper that relies solely on temporal spectral modulation. Because each column in $\Phi$ as shown in Fig.~\ref{fig:theory}(a) is associated with a single wavelength, the full spectral bandwidth remains intact as long as the \textit{two}-dimensional spatio-temporal spectral amplitude modulation does not eliminate any whole column. Instead, amplitude modulation here leads to a change in the axial distribution of each wavelength. In contrast, in traditional pulse shaping \cite{Weiner09Book}, the two-dimensional distribution in Fig.~\ref{fig:theory}(a) is collapsed to one dimension, so that amplitude modulation would inevitably eliminate wavelengths altogether from the spectrum.

Choosing $x'$ for the transverse coordinate in the plane of the SLM and $x$ in the observation plane at axial position $z$ [Fig.~\ref{fig:schematic}(b)], the freely propagating field (the cylindrical lens L$_{1-y}$ affects the field only along $y$) is given by the usual Fresnel integral:
\begin{equation}
E(x,z;\omega)\propto e^{ikz}\int\!dx'\,E(x';\omega)\,e^{i\tfrac{k}{2z}(x-x')^{2}},
\end{equation}
where $E(x';\omega)$ is the field at the SLM along a single column (one temporal frequency $\omega$), and is the product of a linear phase $e^{\pm ik_{x}x'}$ imparted by the SLM and two constant amplitude windows centered at $\mp x_{\mathrm{o}}$ of width $W_{\mathrm{o}}$,
\begin{equation}\label{Eq:SLM_Field}
E(x';\omega)=\widetilde{E}(\omega)e^{\pm ik_{x}x'}\mathrm{rect}\left(\tfrac{x'\pm x_{\mathrm{o}}(\omega)}{W_{\mathrm{o}}(\omega)}\right);
\end{equation}
here both $x_{\mathrm{o}}$ and $W_{\mathrm{o}}$ change with $\omega$, $\mathrm{rect}\left(\tfrac{x-x_{\mathrm{o}}}{W_{\mathrm{o}}}\right)$ is a rectangular (constant) function of width $W_{\mathrm{o}}$ centered at $x\!=\!x_{\mathrm{o}}$, and $\widetilde{E}(\omega)$ is the spectral amplitude for the temporal frequency $\omega$; in our experiments $\widetilde{E}(\omega)$ is approximately constant independently of $\omega$. It is understood throughout that $\omega\!=\!\omega(k_{x};\theta)$, which is obtained by inverting Eq.~\ref{Eq:kxOmega}. From this formulation, we can immediately make two conclusions. First, the field associated with the frequency $\omega$ has its on-axis $x\!=\!0$ peak at
\begin{equation}\label{Eq:AxialLocation}
z(\omega)\!\sim\!\frac{k}{k_{x}}x_{\mathrm{o}}(\omega).
\end{equation}
Second, the transverse width over which the field is spread at $z(\omega)$ is
\begin{equation}\label{fig:AxialWidth}
W(z,\omega)\!\sim\!W_{\mathrm{o}}(\omega)\sqrt{1+\left(\frac{z}{z_{\mathrm{o}}(\omega)}\right)^{2}},
\end{equation}
where $z_{\mathrm{o}}(\omega)\!=\!\pi W_{\mathrm{o}}^{2}(\omega)/\lambda$. From these two observations, we can \textit{reverse engineer} the amplitude mask at the SLM plane by determining $x_{\mathrm{o}}(\omega)$ and $W_{\mathrm{o}}(\omega)$ required to localize the temporal frequency $\omega$ at the specific location or locations along $z$.

By applying this algorithm to all the frequencies $\omega$ modulated by the SLM, thereby sculpting the amplitude mask $A$, we can realize an arbitrary axial spectral encoding for the ST wave packet over any desired bandwidth and along a specified axial range $z$. The spatio-temporal profile of the ST wave packet endowed with the desired axial spectral encoding is thus obtained after integrating over the temporal bandwidth:
\begin{equation}
E(x,z;t)=\int\!d\omega\,E(x,z;\omega)e^{-i\omega(k_{x};\theta)t}.
\end{equation}
After implementing the amplitude mask shown in Fig.~\ref{fig:theory}(a), the time-averaged intensity $I(x,z)\!=\!\int\!dt|E(x,z;t)|^{2}$ remains diffraction-free for an extended distance, but additional transverse sidelobes appear [Fig.~\ref{fig:theory}(b)]. These sidelobes are a consequence of the rectangular-shaped amplitude filter at each temporal frequency; applying a Gaussian amplitude modulation can smooth out these sidelobes.

The axial spectral encoding can be ascertained by examining the evolution of the spectrum along the propagation axis. The axial spectrum $I(x,z;\omega)\!=\!|E(x,z;\omega)|^{2}$ is integrated over the transverse extent shown in Fig.~\ref{fig:theory}(b), from $x\!=\!-200$~$\mu$m to $x\!=\!200$~$\mu$m (corresponding to the diameter of the multimode fiber used in our experiments to collect the spectrum), to yield $I(z,\omega)\!=\!\int\!dxI(x,z;\omega)$, which we plot in Fig.~\ref{fig:theory}(c). For propagation-invariant ST wave packets produced via phase-only modulation, the spectrum remains mostly constant up to $z\!=\!50$~mm. The short-wavelength end of the spectrum gradually diminishes beyond $z\!=\!50$~mm [Fig.~\ref{fig:theory}(c)], which accompanies departure from the diffraction-free condition [Fig.~\ref{fig:theory}(b)]. The ST wave packet simulated in Fig.~\ref{fig:theory} is superluminal ($\theta\!>\!45^{\circ}$, $\widetilde{v}\!>\!c$) \cite{Yessenov19PRA}. Selecting a subluminal ST wave packet instead ($\theta\!<\!45^{\circ}$, $\widetilde{v}\!<\!c$) would result in a diminishing of the long-wavelength portion of its spectrum at the end of the propagation distance.

In contrast, the axial spectral evolution $I(z,\omega)$ for the ST wave packet after amplitude-masking clearly shows an axial spectral encoding with propagation [Fig.~\ref{fig:theory}(c)]: the spectrum at any axial plane is narrow compared to that of the propagation-invariant ST wave packet, and it undergoes a significant red-shift along $z$. This behavior is further highlighted by comparing the spectrum at selected axial planes for both wave packets in Fig.~\ref{fig:theory}(d). In the case of the axially spectral-encoded ST wave packet, the spectral shift along $z$ results in disjoint spectra at the axial start and end points.

\section{Experiment}

We have carried out a sequence of experiments that demonstrate multiple axial spectral encodings for ST wave packets through amplitude-masked phase-modulation of the spatio-temporal spectrum. In each experiment, we measure the axial intensity profile $I(x,z)$, and the axial spectral evolution $I(z,\lambda)$ integrated along $x$ over a fixed transverse width $\sim400$~$\mu$m (as determined by the diameter multimode fiber collecting the signal).

\subsection{Synthesizing axially red-shifting spectra}

As an initial proof-of-concept, we implemented the amplitude-masked phase distributions depicted in Fig.~\ref{fig:redshift}(a) corresponding to two \textit{red}-shifting spectra (the spectrum shifts from shorter to longer wavelengths along $z$). The spectral tilt angle here is $\theta\!=\!50^{\circ}$, and the two amplitude masks $A_{1}$ and $A_{2}$ are designed to yield different axial red-shifting rates. Whereas the width of the amplitude window is a constant $W_{\mathrm{o}}(\lambda)\!=\!W_{\mathrm{o}}$, the center of the window moves along a linear trajectory creating a wedge-shaped mask: $x_{\mathrm{o}}(\lambda)\!=\!x_{\mathrm{max}}\tfrac{\lambda-\lambda_{1}}{\lambda_{2}-\lambda_{1}}$.

In the first example [$A_{1}$ in Fig.~\ref{fig:redshift}, left column], we set $\lambda_{1}\!\approx\!797.9$~nm and $\lambda_{2}\!\approx\!800.1$~nm, and select $W_{\mathrm{o}}$ to produce a measured average on-axis FWHM spectral bandwidth of $\Delta\lambda\sim0.4$~nm. As noted earlier, the measured spectrum when utilizing a multimode fiber to couple light into the OSA is broadened by $\approx\!30\%$ in all cases in comparison with that obtained by using a single-mode fiber. The results reported here are thus a conservative underestimate of the spectral tuning capability of this strategy. The overall shift of the center wavelength of the axial spectral encoding is $\approx 1.3$~nm over a distance of $\approx75$~mm, resulting in an axial red-shifting rate of $\approx17$~pm/mm. This is only a first-order estimate because the axial spectral encoding is not linear along $z$. A linear axial encoding requires implementing a different amplitude mask (see confirmation of this point below). The time-averaged intensity $I(x,z)$ is provided in the Supplementary Material, and is in excellent agreement with the simulation in Fig.~\ref{fig:theory}(b).

In the second example [$A_{2}$ in Fig.~\ref{fig:redshift}, right column], we reduce the axial red-shifting rate to $\approx 11$~pm/mm by selecting a sharper slope of the amplitude mask whereupon $\lambda_{1}\!\approx\!798.3$~nm, $\lambda_{2}\!\approx\!799.7$~nm, while maintaining the same value for $W_{\mathrm{o}}$. In this case, the shift in the center wavelength of the axial spectral encoding is $\approx0.8$~nm observed over a distance of $\approx75$~mm. The spectra at selected axial planes are plotted for these two examples in Fig.~\ref{fig:redshift}(c). As a point of reference, we provide a Supplementary Movie (Movie10) for the spectral evolution of the ST wave packet with the amplitude mask removed, which shows a mostly invariant axial spectrum. Supplementary Movies corresponding to the amplitude masks $A_{1}$ and $A_{2}$ (Movie1 and Movie2, respectively) display the change in the spectral encoding along $z$. Together, these two examples provide a demonstration of the general control strategy over the rate of axial spectral shift.

\subsection{Synthesizing axially blue-shifting spectra}

\begin{figure}[t!]
\centering
\includegraphics[width=8.8cm]{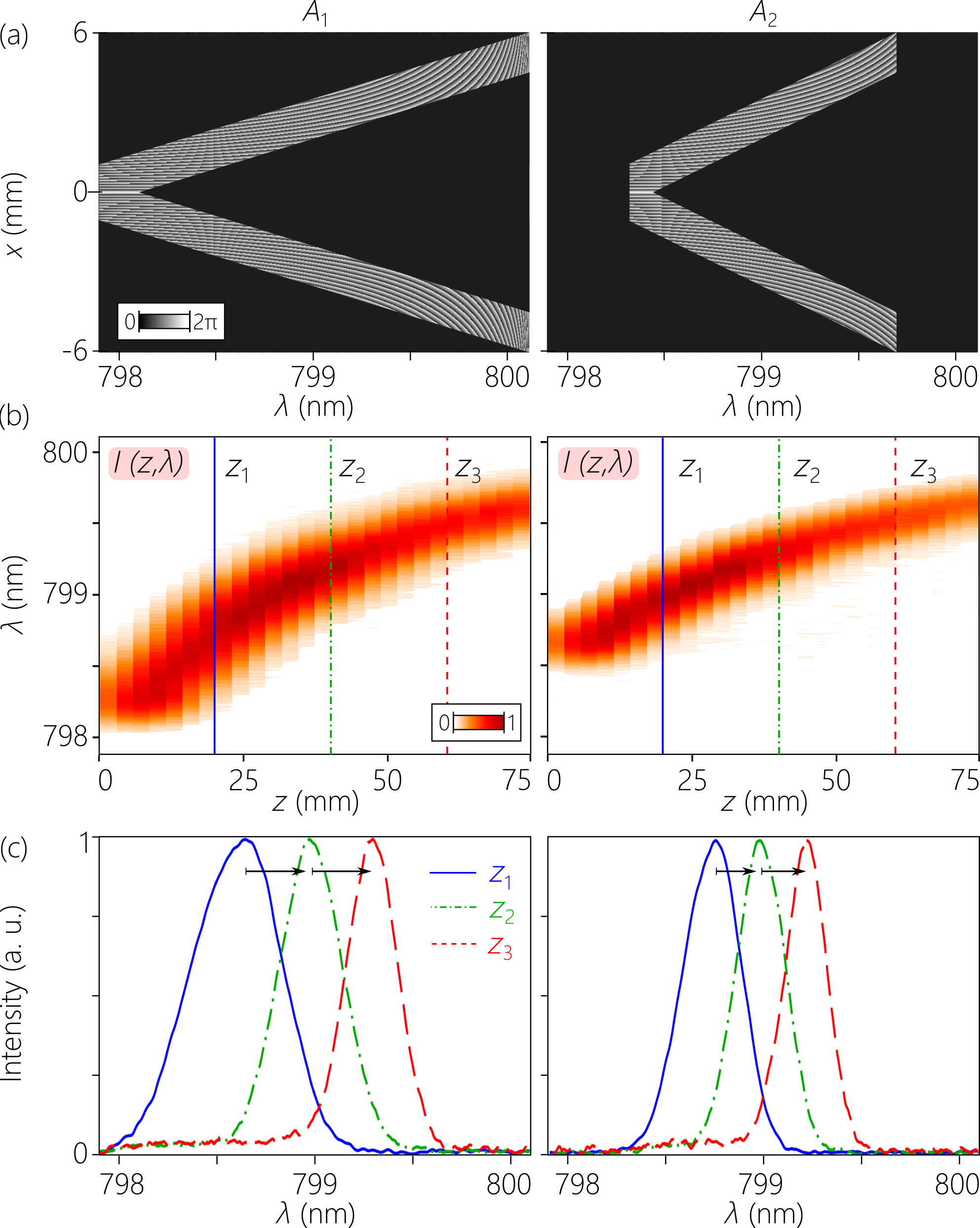}
\caption{(a) Two amplitude-masked phase distributions, $A_{1}$ and $A_{2}$ (the underlying phase distribution corresponds to a spectral tilt angle of $\theta\!=\!50^{\circ}$), designed to produce axially \textit{red}-shifting spectra by changing the the slope of the `wedge' profile. (b) Measured axial spectra $I(z,\lambda)$. (c) Measured spectra in selected axial planes. Black arrows indicate the direction of the spectral shift. See Supplementary Movie1 and Movie2.}
\label{fig:redshift}
\end{figure}

\begin{figure}[t!]
\centering
\includegraphics[width=8.6cm]{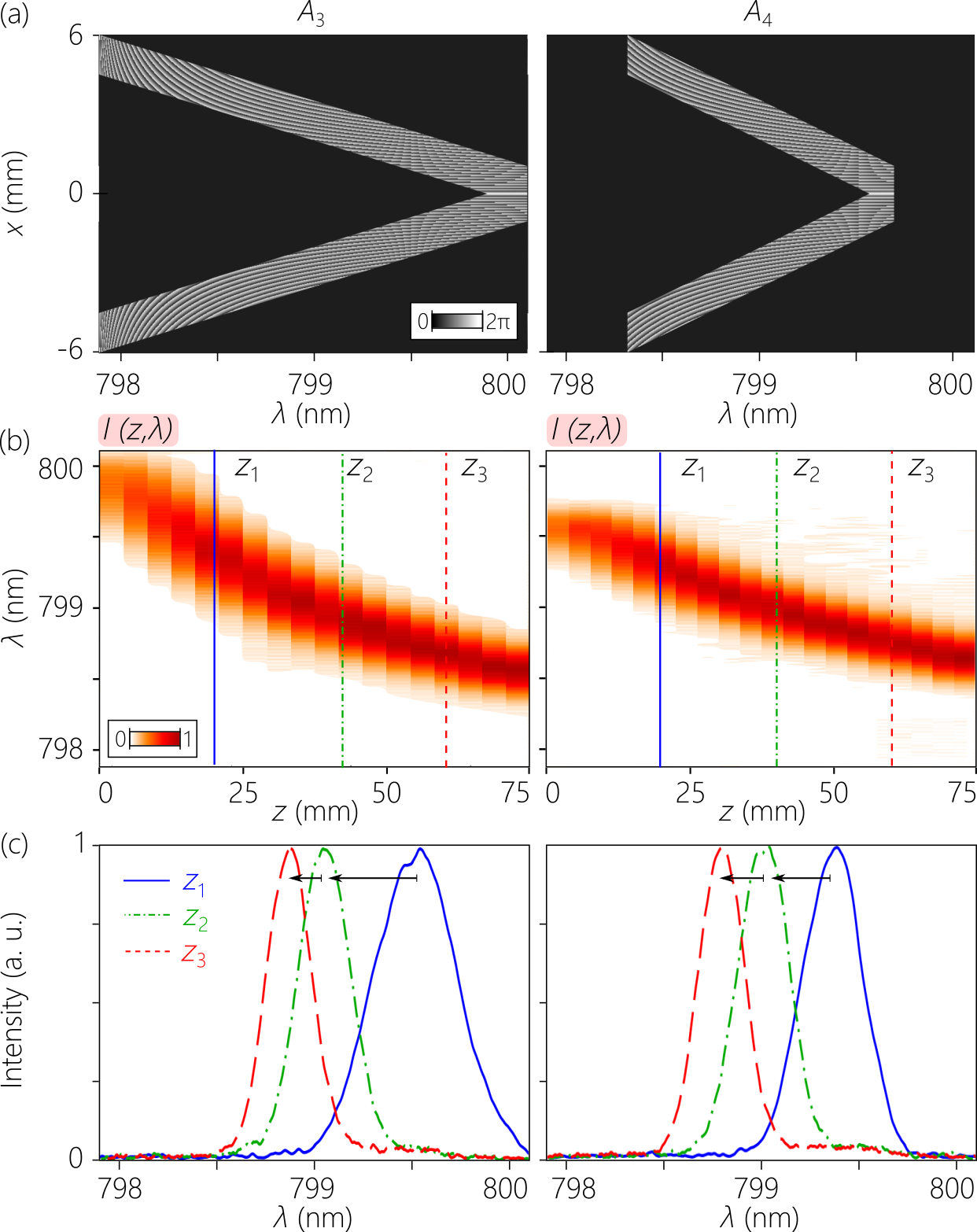}
\caption{(a) Two amplitude-masked phase distributions, $A_{3}$ and $A_{4}$ (the underlying phase distribution corresponds to a spectral tilt angle of $\theta\!=\!40^{\circ}$) designed to produce axially \textit{blue}-shifting spectra by changing the the slope of the `wedge' profile. (b) Measured axial spectra $I(z,\lambda)$. (c) Measured spectra in selected axial planes. Black arrows indicate the direction of the spectral shift. See Supplementary Movie3 and Movie4.}
\label{fig:blueshift}
\end{figure}

Next we demonstrate that the same approach is applicable to \textit{blue}-shifting axial spectral encoding (the spectrum shifts from longer to shorter wavelengths along $z$). The designed amplitude masks for two different axial blue-shifting rates are shown in Fig.~\ref{fig:blueshift}(a). The slope of the wedge-like amplitude mask has the opposite sign of the red-shifting masks in Fig.~\ref{fig:redshift}(a). Here the underlying phase distribution corresponds to a subluminal spectral tilt angle of $\theta\!=\!40^{\circ}$. In the first example ($A_{3}$ in Fig.~\ref{fig:blueshift},left column) a blue-shifting spectral rate of $\approx-18$~pm/mm is achieved, corresponding to a total spectral shift of $\Delta\lambda\!\approx\!1.3$~nm over $\sim75$~mm; and in the second example ($A_{4}$ in Fig.~\ref{fig:blueshift}, right column) $\approx-12$~pm/mm, corresponding to $\Delta\lambda\!\approx\!0.9$~nm over $\sim75$~mm; see Fig.~\ref{fig:blueshift}(b,c). These values approximately mirror the rates observed in the red-shifting scenario in Fig.~\ref{fig:redshift}. The measured time-averaged intensity $I(x,z)$ is given in the Supplementary Material. Supplementary Movies corresponding to the amplitude masks $A_{3}$ and $A_{4}$ (Movie3 and Movie4, respectively) display the measured change in the spectral encoding along $z$.

\begin{figure}[t!]
\centering
\includegraphics[width=8.6cm]{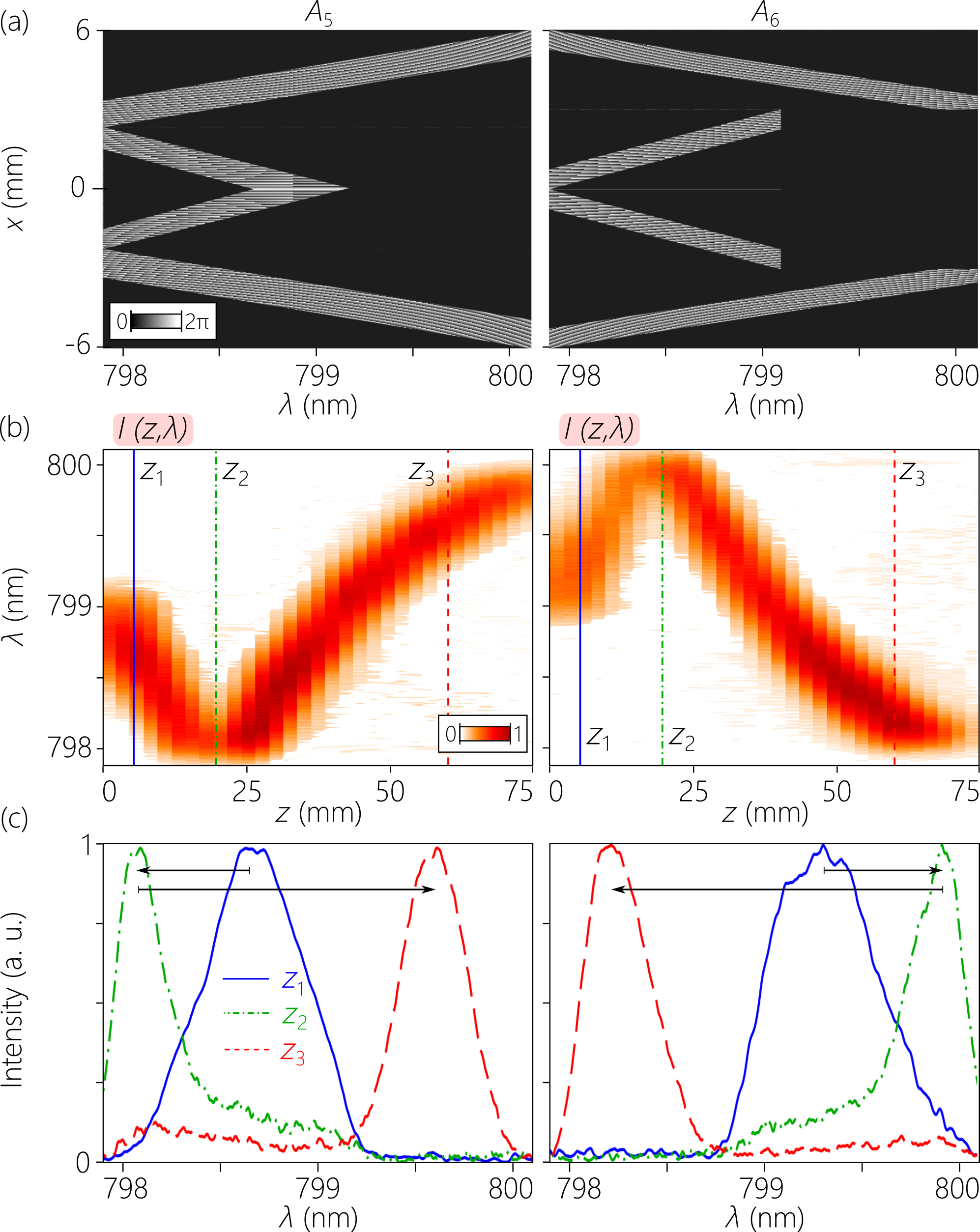}
\caption{(a) Two amplitude-masked phase distributions, $A_{5}$ and $A_{6}$ (corresponding to spectral tilt angles of $\theta\!=\!50^{\circ}$ and $\theta\!=\!40^{\circ}$, respectively). The masked distribution $A_{5}$ on the left produces a ST wave packet with blue-shifting followed by red-shifting axial spectral encoding; whereas the masked distribution $A_{6}$ on the right produces a ST wave packet with red-shifting followed by blue-shifting axial spectral encoding. (b) Measured axial spectra $I(z,\lambda)$. (c) Measured spectra in selected axial planes. Black arrows indicate the direction of the spectral shift. See Supplementary Movie5 and Movie6.}
\label{fig:bishift}
\end{figure}

\subsection{Synthesizing axially bidirectional-shifting spectra}

The axial spectral encodings in the previous subsections all feature monotonic (or uni-directional) spectral shifts. However, our approach is more general and, in principle, can establish arbitrary evolution of the on-axis spectrum. In Fig.~\ref{fig:bishift}(a), we show the amplitude-masked phase distributions required to produce axial spectral encodings that feature nonmonotonic (here, bidirectional) spectral shifts. In one example ($A_{5}$ in Fig.~\ref{fig:bishift}(a), left column), the amplitude mask is designed to result initially in a blue-shift (from $\approx799$~nm to $\approx798$~nm over $\sim20$~mm) followed by a red-shift (from $\approx798$~nm to $\approx800$~nm over $\sim50$~mm). The measured axial spectral evolution $I(z,\lambda)$ is plotted in Fig.~\ref{fig:bishift}(b) along with the spectra at selected axial planes. In the second example ($A_{6}$ in Fig.~\ref{fig:bishift}(a), right column), the amplitude mask is designed to result initially in a red-shift (from $\approx799$~nm to $\approx800$~nm over $\sim20$~mm) followed by a blue-shift (from $\approx800$~nm to $\approx798$~nm over $\sim50$~mm); see Fig.~\ref{fig:bishift}(b,c) for the measured $I(z,\lambda)$. Supplementary Movies corresponding to the amplitude masks $A_{5}$ and $A_{6}$ (Movie5 and Movie6, respectively) display the change in the spectral encoding along $z$.

\subsection{Synthesizing an axially arbitrary-shifting spectrum}

\begin{figure}[t!]
\centering
\includegraphics[width=8.6cm]{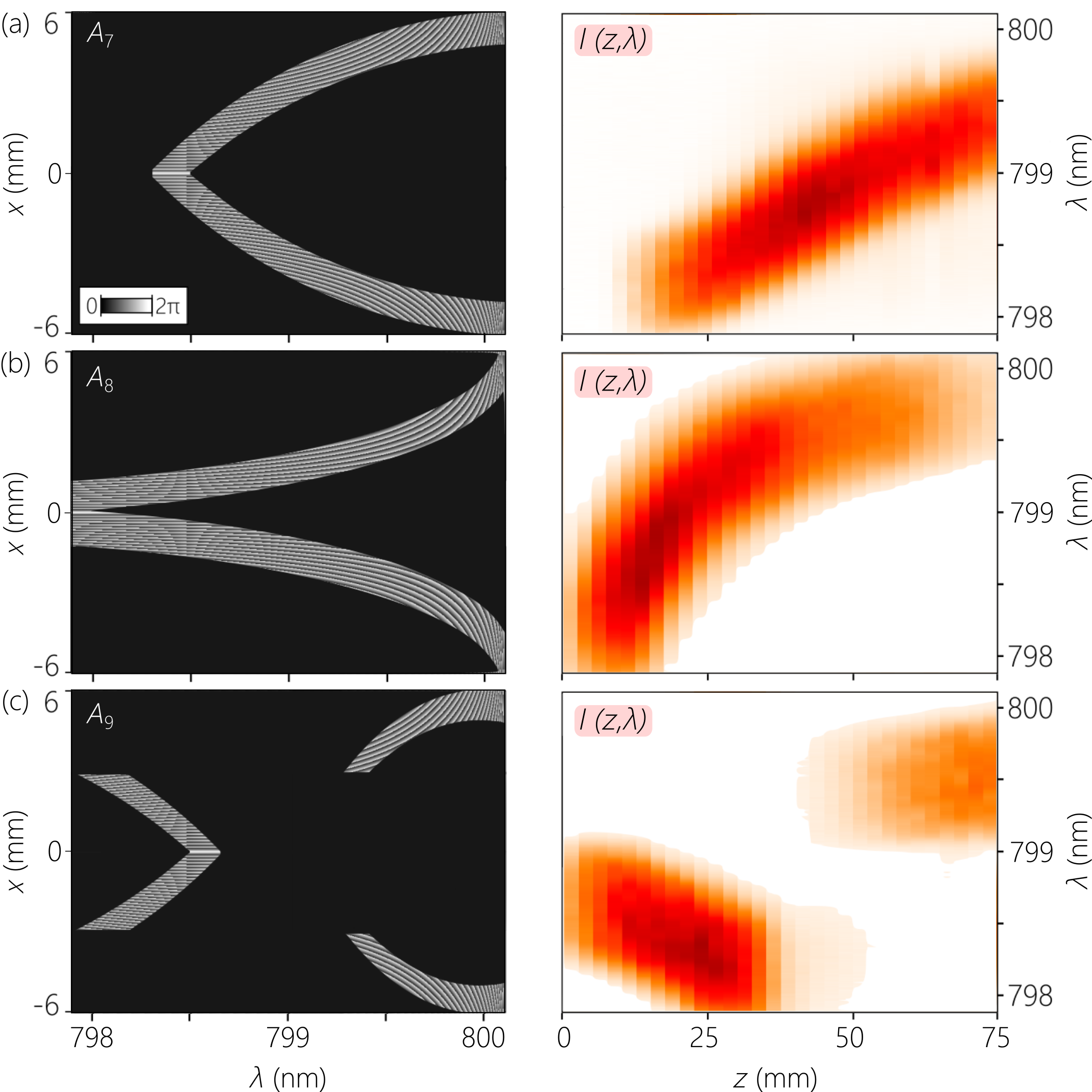}
\caption{(a) An amplitude-masked phase distribution shaped with a supralinear curvature $A_{7}$ produces a \textit{linear} shift in the central wavelength along $z$. (b) An amplitude-masked phase distribution shaped with a sublinear curvature $A_{8}$ leads to a higher \textit{acceleration} rate of the axial spectral encoding along $z$ with respect to that shown in Fig.~\ref{fig:redshift}. (c) A disjointed amplitude masked phase modulation $A_{9}$ yields a disjointed axial spectral encoding. See Supplementary Movie7, Movie8, and Movie9.}
\label{fig:misc}
\end{figure}

In the three classes of axial spectral encoding provided above (red-, blue-, and bidirectional-shifting), the amplitude-masked phase distributions implemented by the SLM in Fig.~\ref{fig:schematic}(b) feature wedge-like masks with a linear slope for $x_{\mathrm{o}}(\lambda)$. As seen in Fig.~\ref{fig:redshift}-Fig.~\ref{fig:bishift}, such amplitude masks result in a nonlinear trajectory of the axial spectral evolution. 

In order to exercise precise control over the axial spectral encoding, we construct a predictive theory for its quantitative dependence on the amplitude mask shape. To enforce a specific axial spectral encoding characterized by an on-axis spectral profile $z(\lambda)$ over a particular stretch of the $z$-axis, Eqs.~\ref{Eq:kxOmega}, \ref{Eq:SLM_Field}, and \ref{Eq:AxialLocation} allow us to establish the following relationships for the amplitude-mask:
\begin{eqnarray}\label{Eq:Trajectory}
x_{\mathrm{o}}(\lambda)&=&z(\lambda)\sqrt{1-\cot{\theta}}\frac{\lambda}{\lambda_{\mathrm{o}}}\sqrt{\frac{\lambda_{\mathrm{o}}}{\lambda}-1},\,\,\theta>45^{\circ}\nonumber\\
x_{\mathrm{o}}(\lambda)&=&z(\lambda)\sqrt{\cot{\theta}-1}\frac{\lambda}{\lambda_{\mathrm{o}}}\sqrt{1-\frac{\lambda_{\mathrm{o}}}{\lambda}},\,\,\theta<45^{\circ}.
\end{eqnarray}
These equations provide a one-to-one correspondence between the position of the amplitude modulation and the axially realized wavelength. Making use of these equations, we realize the sophisticated axial spectral encodings plotted in Fig.~\ref{fig:misc}. First, we demonstrate the linear axial spectral encoding plotted in Fig.~\ref{fig:misc}(a), in contrast to those in Figs.\ref{fig:redshift}-\ref{fig:bishift}, using an amplitude mask $A_{7}$ that follows a supra-linear pattern rather than a linear one. Second, we \textit{accelerate} the axial spectral encoding in Fig.~\ref{fig:misc}(b) by relying on an amplitude mask $A_{8}$ with a sub-linear pattern. Finally, we show in Fig.~\ref{fig:misc}(c) an example of an axial spectral encoding featuring a spectral discontinuity comprising a blue-shifting axial section, followed by a step to a red-shifting axial section, with a spectral gap separating the two sections produced by the amplitude mask $A_{9}$. We provide in the Supplementary Materials the exact equations for the amplitude masks shown in Fig.~\ref{fig:misc}. Supplementary Movies corresponding to the amplitude masks $A_{7}$, $A_{8}$, and $A_{9}$ (Movie7, Movie8, and Movie9 respectively) display the change in the spectral encoding along $z$. 

\begin{figure}[t!]
\centering
\includegraphics[width=8.6cm]{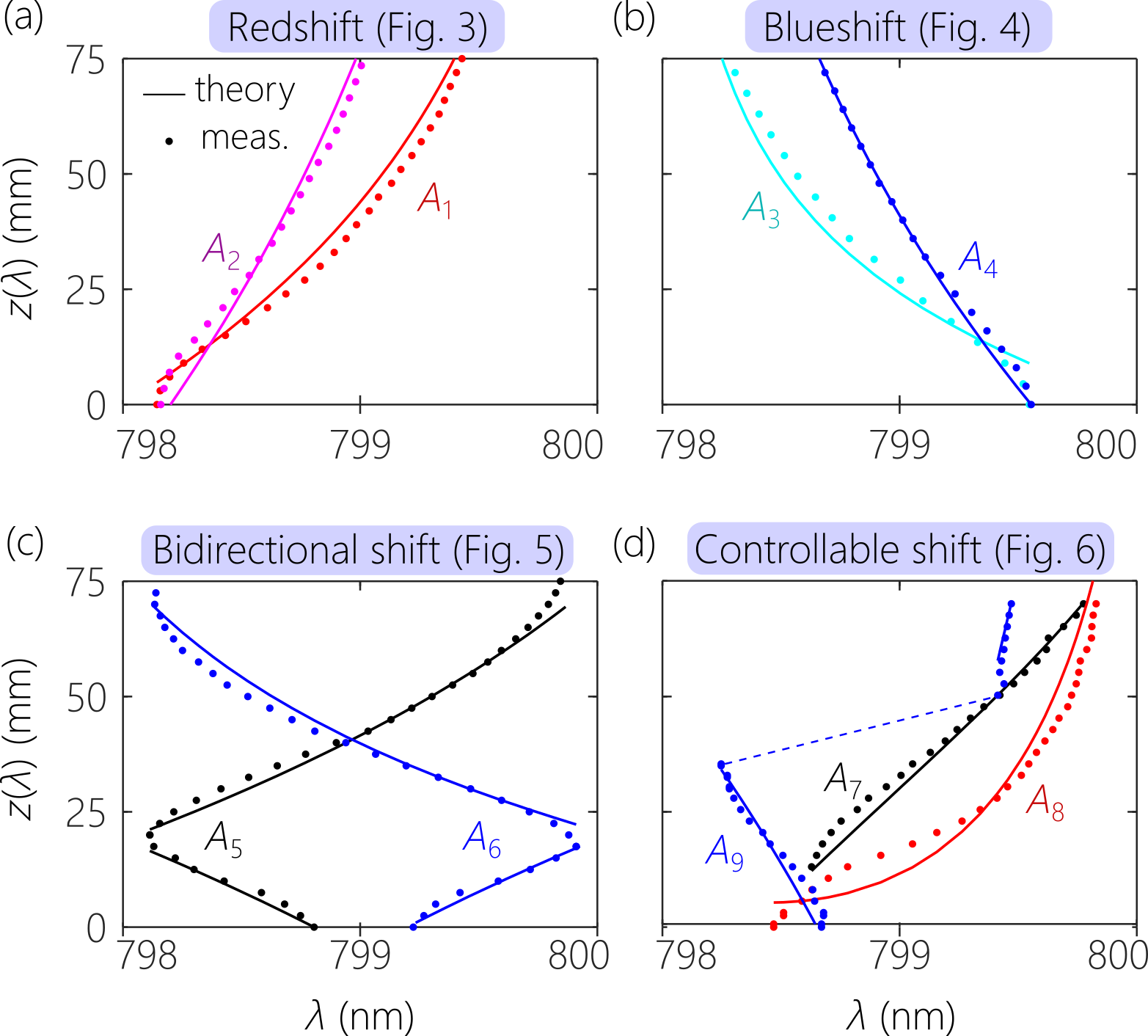}
\caption{Summary of the axial dynamics along $z$ of the center of the spectrum for axial spectrally encoded ST wave packets. (a) Red-shifting and (b) blue-shifting ST wave packets from Fig.~\ref{fig:redshift} and Fig.~\ref{fig:blueshift}, respectively. (c) Bidirectional axially spectral-encoded ST wave packets from Fig.~\ref{fig:bishift}. (d) Controllable axial spectral encodings corresponding to Fig.~\ref{fig:misc}. In each case, the data corresponds to the evolution of the central wavelength of a Gaussian distribution fitted to the spectra in Figs.~\ref{fig:redshift}-\ref{fig:misc}. The solid curves are theoretical predictions based on Eq.~\ref{Eq:Trajectory}.}
\label{fig:lineouts}
\end{figure}

We summarize in Fig.~\ref{fig:lineouts} the axial spectral encodings demonstrated in Figs.~\ref{fig:redshift}-\ref{fig:misc}: red-shifting spectra in Fig.~\ref{fig:lineouts}(a), blue-shifting spectra in Fig.~\ref{fig:lineouts}(b), bidirectional spectra in Fig.~\ref{fig:lineouts}(c), and the precisely controlled spectral encodings from Fig.~\ref{fig:misc} in Fig.~\ref{fig:lineouts}(d). In each case, we plot the data in the form of the \textit{center} of a Gaussian distribution fitted to the spectrum at each axial position $z$. With each data set, we plot the axial spectral encoding predicted based on Eq.~\ref{Eq:Trajectory} (see Supplementary Materials). We find excellent agreement between the data and the predictions in all cases.

\section{Time-Resolved Measurements}

We have so far presented time-averaged spectral measurements of the evolution of the on-axis spectrum for the synthesized ST wave packets. In this Section, we present \textit{time-resolved} measurements of their spatio-temporal profiles in comparison with those of propagation-invariant ST wave packets. We aim from these measurements to show that the amplitude-masking of the phase modulation does not impact the wave packet group velocity.

\begin{figure}[t!]
\centering
\includegraphics[width=8.6cm]{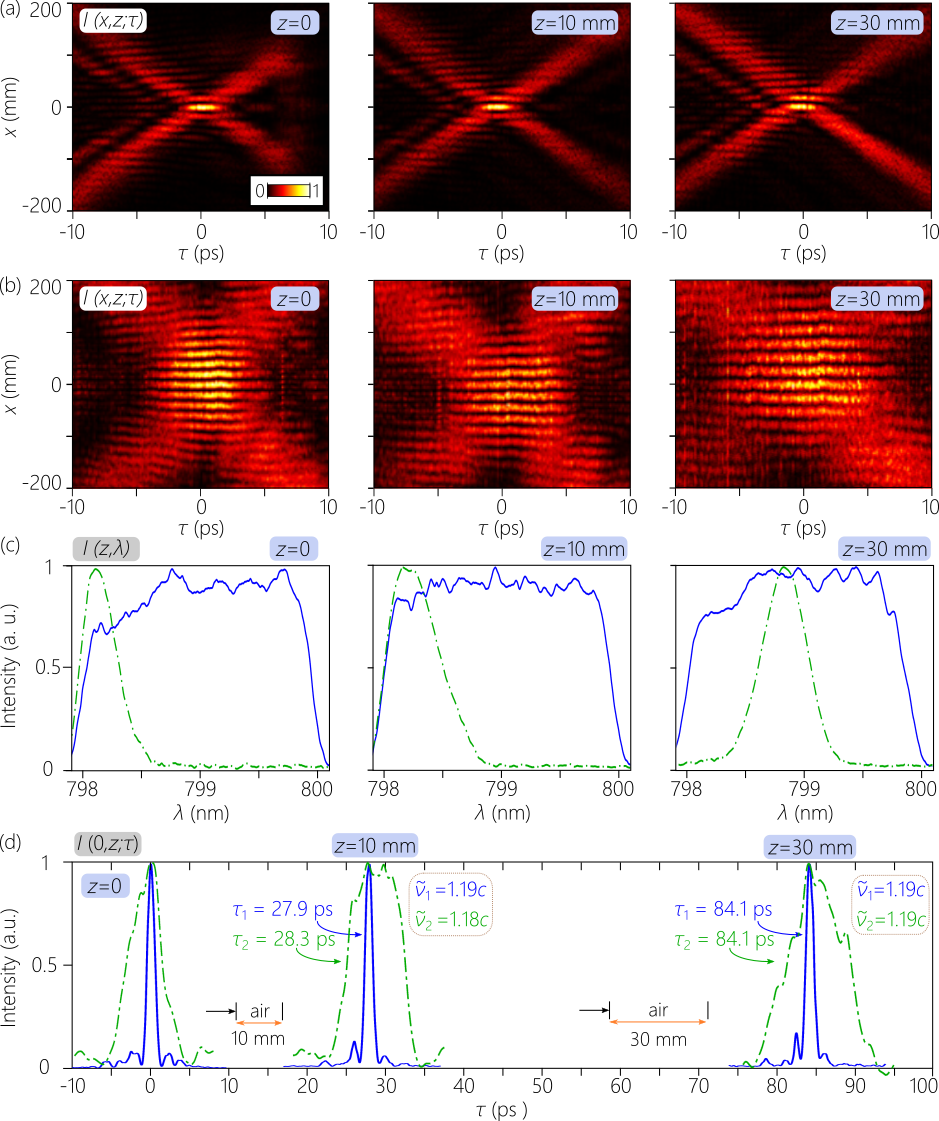}
\caption{(a) The spatio-temporal intensity profile $I(x,z;\tau)$ measured at $z\!=\!0$, 10, and 30~mm for a propagation-invariant ST wave packet produced by a phase-only spatio-temporal spectral modulation, and (b) for an axially spectral-encoded ST wave packet produced using the same phase distribution with $\theta\!=\!50^{\circ}$ after amplitude-masking [Fig~\ref{fig:TRdata}(a), left column]. (c) The measured spectra $I(z,\lambda)$ measured at $z\!=\!0,10$ and $30$ mm for the ST wave packets in (a) and (b). The broad spectrum of the propagation-invariant ST wave packet (continuous curves) does not change with $z$, whereas the narrow spectrum of the axially spectral-encoded ST wave packet undergoes a red-shift with $z$. (d) Normalized temporal pulse profiles $I(x\!=\!0,z;\tau)$ at the center of the transverse profiles of the ST wave packets in (a) and (b) at  $z\!=\!0$, 10, and 30~mm. The two wave packets travel at the same group velocity.}
\label{fig:TRdata}
\end{figure}

We determined the group velocity of a ST wave packet by measuring its relative group delay with respect to a reference pulse traveling at $c$; see \cite{Kondakci19NC,Bhaduri19Optica,Yessenov19OE}. We plot in Fig.~\ref{fig:TRdata}(a) the spatio-temporal profiles $I(x,z;\tau)$ at three axial planes ($z\!=\!0$, 10, and 30~mm) for two ST wave packets both having $\theta\!=\!50^{\circ}$ ($\widetilde{v}\!=\!c\tan{50^{\circ}}\!\approx\!1.19c$). One ST wave packet was synthesized via phase-only spectral modulation and is thus propagation invariant [Fig.~\ref{fig:TRdata}(a)], and the other via the amplitude-masked phase modulation from Fig.~\ref{fig:redshift}(a) that produces a red-shifting axial spectral encoding [Fig.~\ref{fig:TRdata}(b)]. The width of the wave packet at the center of its spatial profile $x\!=\!0$ is smaller for the propagation-invariant ST wave packet in comparison with its axial spectrally encoded counterpart, as expected because of its broader spectrum. 

We plot in Fig.~\ref{fig:TRdata}(c) the spectrum $I(z,\lambda)$ at the three axial planes $z\!=\!0$, 10, and 30~mm. The 2-nm-bandwidth spectra for the propagation-invariant ST wave packet are unchanged with $z$, whereas the narrower spectra for the axially spectral-encoded ST wave packet red-shifts with $z$. The narrower spectrum of the latter results in a broad temporal profile [Fig.~\ref{fig:TRdata}(b)]; whereas the broad spectrum of the propagation-invariant ST wave packet results in a narrow temporal profile [Fig.~\ref{fig:TRdata}(a)]. In both cases however, the spatio-temporal profile is preserved over the propagation distance.

In Fig.~\ref{fig:TRdata}(d) we plot the ST wave packet temporal profiles at the beam center $I(0,z;\tau)$ measured at $z\!=\!0$, 10, and 30~mm. The initial measurement at $z\!=\!0$ is the reference point for measuring subsequent group delays. Crucially, the group delays incurred by both wave packets are equal at the two planes $z\!=\!10$ and 30~mm, thereby proving that amplitude-masking does \textit{not} impact the group velocity. At $z\!=\!10$~mm the group delay is $\tau\!\approx\!28$~ps and at $z\!=\!30$~mm is $\tau\!\approx\!84$~ps, which are consistent with a group velocity of $\widetilde{v}\!\approx\!1.19c\pm0.018c$ (note that this superluminal group velocity does not contradict relativistic causality \cite{Shaarawi95JMP,SaariPRA18,Saari19PRA,Yessenov19OE}).

Finally, besides confirming that amplitude-masking does not impact the group velocity of the ST wave packet, we have also verified that the \textit{same} axial spectral encoding can be implemented using ST wave packets of \textit{different} group velocity. We present the data in the Supplementary Materials measurements of the axial spectral evolution of wave packets with $\theta\!=\!50^{\circ}$ ($\widetilde{v}\!\approx\!1.19c$) and $\theta\!=\!70^{\circ}$ ($\widetilde{v}\!\approx\!2.75c$) using the same amplitude mask that induces a red-shifting axial spectrum. 

\section{Discussion and Conclusion}

The added feature of axial spectral encoding demonstrated here with ST wave packets can now be combined with previous achievements made utilizing these unique pulsed beams, such as long-distance propagation \cite{Bhaduri18OE,Bhaduri19OL} and self-healing \cite{Kondakci18OL}. One can now potentially rely on the wavelength of light reflected from an object as an axial-ranging stamp; that is, the wavelength of the reflected or scattered light will correlate with the object's position and thus indicate its distance from the source. Because ST wave packets can be synthesized with narrow transverse widths ($\sim7$~$\mu$m in \cite{Kondakci17NP} and $\sim4$~$\mu$m in \cite{Shiri20OL}) and large depths of focus (25~mm for the 7-$\mu$m-wide wave packet in \cite{Kondakci17NP}), they can be useful in the context of optical microscopy. Alternatively, the demonstration of propagation lengths approaching 100~m with ST wave packets \cite{Bhaduri19OL} suggests the possibility of implementing such a protocol in remote sensing. Moreover, the experiments reported here for axial spectral encoding all made use of coherent ultrafast laser pulses. However, such effects can also be achieved with incoherent light from a LED \cite{Yessenov19Optica} or superluminescent LED \cite{Yessenov19OL}.
 
We are now in a position to compare the two major strategies for creating optical wave packets of controllable group velocity in free space: the flying-focus approach \cite{SaintMarie17Optica,Froula18NP,Jolly20OE} and ST wave packets. The flying-focus strategy produces wave packets that are \textit{not} diffraction-free but whose spectrum evolves monotonically along the propagation axis, and this spectral evolution \textit{determines} the group velocity of the wave packet. The spectral evolution and the group velocity are \textit{not} independently addressable in this case. On the other hand, ST wave packets are diffraction-free and have controllable group velocities that can be tuned by tailoring their spatio-temporal structure via spectral-phase-only modulation. The spectrum of such a wave packet is also propagation-invariant. However, we have shown here that by exploiting an amplitude-masked spectral-phase modulation the on-axis spectrum can be made to evolve with arbitrary axial profiles, whether monotonic or non-monotonic, independently of the group velocity while maintaining the propagation-invariance of the wave packet.

In conclusion, we have presented a comprehensive scheme for controlling the evolution of the on-axis spectrum of a freely propagating ST wave packet: the group velocity \textit{and} the axial spectral evolution are arbitrary and tuned independently, while maintaining propagation invariance. Using this approach, we have produced axial spectral encodings in the form of red-shifting, blue-shifting, bidirectional spectral shifting (blue-shifting followed with red-shifting or vice versa over prescribed axial distances), and controlled rates of spectral shifts (linear or accelerating). This scheme is equally applicable to coherent ultrafast pulses and to broadband incoherent stationary fields. Unlike previous efforts on ST wave packets over the past $\sim40$~years that have been devoted to maintaining their propagation invariance, we have demonstrated here that one degree of freedom (the on-axis spectrum) can be isolated and control of its axial propagation exercised while retaining the propagation-invariance of its other degrees of freedom. These results may be useful in spectral stamping for axial range-finding in microscopy and remote-sensing.

\vspace{2mm}\noindent\textbf{Acknowledgments.} We thank G.~Li and P.~J.~Delfyett for loan of equipment. This work was funded by the U.S. Office of Naval Research (ONR) contracts N00014-17-1-2458 and N00014-19-1-2192, and ONR MURI contract N00014-20-1-2789.

\bibliography{diffraction}

\end{document}